\documentclass[aps,prb,twocolumn,amsfonts,amsmath,amssymb,superscriptaddress]{revtex4}
\usepackage{graphicx}
\usepackage{dcolumn}
\usepackage{bm}

\usepackage{colordvi}
\usepackage{color}
\begin{document}

\title[A direct comparison of CVD-grown and exfoliated MoS$_2$ using optical spectroscopy]{A direct comparison of CVD-grown and exfoliated MoS$_2$ using optical spectroscopy}

\author{G. Plechinger$^1$, J. Mann$^2$,E. Preciado$^2$, D. Barroso$^2$,  A. Nguyen$^2$, J. Eroms$^1$, C.\ Sch{\"u}ller$^1$, L. Bartels$^2$ and T.\ Korn$^{1,*}$}
\address{$^1$Institut f\"ur Experimentelle und Angewandte Physik,
Universit\"at Regensburg, D-93040 Regensburg, Germany}
\address{$^2$Chemistry, Physics, and Materials Science and Engineering, University of California, CA 92521 Riverside, USA}

\begin{abstract}
MoS$_2$ is a highly interesting material system, which exhibits a crossover from an indirect band gap in the bulk crystal to a direct gap for single layers. Here, we perform a direct comparison between large-area MoS$_2$ films grown by chemical vapor deposition (CVD) and MoS$_2$ flakes prepared by mechanical exfoliation from natural bulk crystal. Raman spectroscopy measurements show differences between the in-plane and out-of-plane phonon mode positions in CVD-grown and exfoliated MoS$_2$.  Photoluminescence (PL) mapping reveals large regions in the CVD-grown films that emit strong PL at room temperature, and low-temperature PL scans demonstrate a large spectral shift of the A exciton emission as a function of position. Polarization-resolved PL measurements under near-resonant excitation conditions show a strong circular polarization of the PL, corresponding to a valley polarization.
\end{abstract}
\maketitle
\section{Introduction}
In recent years, two-dimensional crystal structures~\cite{Novoselov26072005} have attracted a lot of attention. Many layered crystals can be mechanically exfoliated to yield flakes of single-layer thickness, just as graphene can be prepared from graphite. Transition metal dichalcogenides such as MoS$_2$ and WS$_2$ are currently studied by many groups, as they show drastic changes of the band structure with reduction of the number of layers. For MoS$_2$, a transition from indirect to direct band gap was theoretically predicted~\cite{Eriksson09} and experimentally observed~\cite{Heinz_PRL10,Splen_Nano10}, making the material interesting for possible opto-electronic applications. The band structure of monolayer MoS$_2$ allows for optical generation and detection of a valley polarization~\cite{Xiao12}. Large excitonic effects are predicted due to the two-dimensional crystal structure~\cite{Rama12}. Additionally, a number of studies demonstrated transistor~\cite{Kis11} and memory devices~\cite{Kis_Memory} with attractive properties. The majority of experimental studies have been performed on MoS$_2$ flakes exfoliated from natural bulk crystal. While this material is readily available for fundamental research and even allows for preparation of integrated circuits~\cite{Wang_Integrated_Nano12}, commercial  application of atomically thin MoS$_2$ will require wafer-scale deposition of films with well-controlled properties. A number of different methods for preparation of large-area MoS$_2$ films have been developed in recent years. It was demonstrated that chemical exfoliation of bulk MoS$_2$, in combination with mild annealing, can yield thin films which show pronounced PL~\cite{Eda11}. Another approach is based on the sulfurization of thin metallic Mo films in Nitrogen~\cite{Zhan12} or Argon~\cite{Duesberg13} atmosphere.  Chemical and physical vapor deposition (CVD/PVD) of MoS$_2$ was reported for different substrates, including graphene~\cite{Shi_CVD}, copper~\cite{Bartels_CVD} and sapphire~\cite{Wu_vaporTransport}.

Here, we report on optical spectroscopy of large-area MoS$_2$ films grown by CVD on SiO$_2$. We utilize Raman spectroscopy, as well as photoluminescence (PL) spectroscopy at room temperature and low temperatures to directly compare the properties of these films to MoS$_2$ flakes prepared by mechanical exfoliation. PL mapping reveals large regions in the CVD-grown films that emit strong PL at room temperature, and low-temperature PL scans demonstrate a large spectral shift of the A exciton emission peak as a function of position. A strong circular polarization of the PL is observed under near-resonant excitation conditions at low temperature, indicating an optically oriented valley polarization. We find pronounced differences between the in-plane and out-of-plane phonon mode positions and linewidths in CVD-grown and exfoliated MoS$_2$, and a clear correlation between the Raman and PL spectra in the CVD-grown films.
\section{Methods}
\subsection{Sample preparation and optical characterization}
\begin{figure}[h]%
\centering
\includegraphics[width= \linewidth]{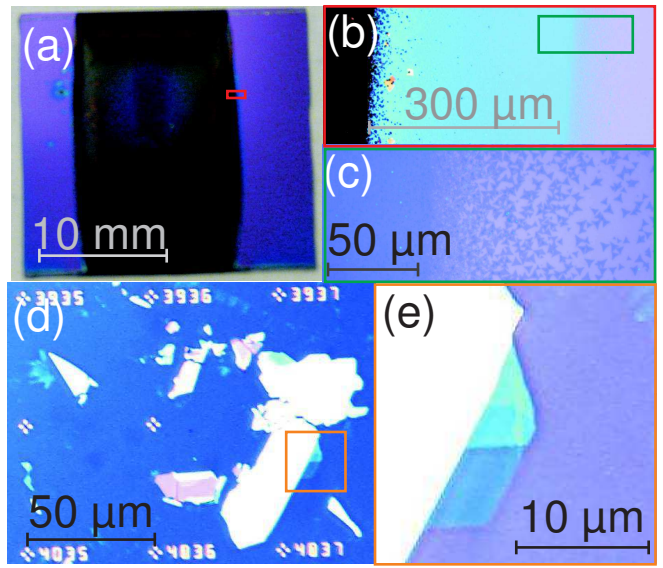}
\caption{(a) Photograph of sample after CVD growth. (b) Optical micrograph of red area indicated in (a) showing monolayer MoS$_2$ region. (c) High-magnification micrograph of green area in (b) showing transition from individual islands to monolayer. (d) Optical micrograph of  MoS$_2$ flakes prepared by mechanical exfoliation on a SiO$_2$/Si wafer. (e) High-magnification micrograph of orange square area in (d). The bottom part of the flake (light blue color) is a monolayer.}
\label{Fig1_samples.eps}
\end{figure}

The synthesis processes for the MoS$_{2}$ growth is as follows. We use two alumina crucibles (Aldrich Z561738, 70~mm $\times$ 14~mm $\times$ 10~mm), one containing .55 g of MoO$_{3}$ (Aldrich 99.5\%) and one containing .70 g of sulfur powder (Alfa 99.5\%) as our Mo and S sources, respectively. These crucibles are placed in a quartz process tube (2~inch diameter), which is inserted into a three zone tube furnace (Mellen TT12), in which only the center zone is heated. A flow of nitrogen gas (99.999\%) is used to purge the tube (5.0 SCFH) for 15 minutes after which the flow is reduced (0.5 SCFH). The crucible containing MoO$_{3}$ is placed at the center of the heated zone with the substrate placed across the middle of the crucible.  The crucible containing sulfur is placed 30 cm away from the crucible containing MoO$_{3}$ in the upstream direction. Our substrate is a square (2~cm width) piece of a boron-doped Si(100) wafer with a 300 nm oxide layer (SUMCO). The substrate is cleaned prior to use by an IPA rinse followed by a piranha etch (3 parts sulfuric acid to 1 part hydrogen peroxide(30\%)) for 1 hr. The center zone of the furnace is heated to 700$^\circ$~C, as measured by a type-K thermocouple at the outer surface of the process tube, in 43 minutes while the sulfur is simultaneously heated to $\sim 200^\circ$~C (due to the temperature gradient from center heated zone) and is held for at this temperature for 10 minutes. Subsequently, the furnace is turned off and allowed to naturally cool to 500$^\circ$~C. Compressed air is then used to rapidly cool the furnace to room temperature. This procedure yields samples as shown in Fig.~\ref{Fig1_samples.eps}(a), where a deposition of material is clearly visible on a 10~mm wide strip along the center of the substrate. Optical microscopy reveals regions of uniform apparent color, indicating a constant MoS$_2$ layer thickness, which are about 300~$\mu$m wide, at the edges of this strip, see Fig.~\ref{Fig1_samples.eps}(b). The long axis of these regions is aligned along the nitrogen flow orientation during growth. We will show below that a pronounced photoluminescence emission corresponding to single-layer MoS$_2$ is observed in these regions. At the edge of these regions, the continuous MoS$_2$ layer transitions to a disordered array of individual islands, which are mostly triangular in shape and show the same apparent color (Fig.~\ref{Fig1_samples.eps}(c)).

The MoS$_2$ flakes used for comparison to the CVD-grown film are prepared using the mechanical exfoliation method well-established for graphene, from natural MoS$_2$. A  p-doped  silicon wafer with 300~nm SiO$_2$ layer and lithographically defined metal markers is used as a substrate. An optical microscope is used to identify flakes that contain single or few layer-regions based on the optical contrast~\cite{castellanos-gomez:213116,Kis11}, the thickness of individual exfoliated flakes is determined with Raman spectroscopy based on the frequency difference of the characteristic Raman modes E$^1_{2g}$ and A$_{1g}$~\cite{Heinz_ACSNano10, Molina11}. The mechanical exfoliation yields a small number of single- and few-layer MoS$_2$ flakes with a size of several micrometers. Fig.~\ref{Fig1_samples.eps}(d) shows a typical optical microscope image of flakes prepared by mechanical exfoliation, the larger-magnification image in Fig.~\ref{Fig1_samples.eps}(e) contains a flake with a single-layer region.
\subsection{Optical spectroscopy}
Raman spectroscopy measurements are performed at room temperature. For this, we utilize a microscope setup, in which a 532~nm cw laser is coupled either into a 100x (1~$\mu$m laser spot size) or a 20x (3~$\mu$m laser spot size) microscope objective, which also collects the scattered light in backscattering geometry. The scattered light is recorded using a  grating spectrometer equipped with a cooled charge-coupled device (CCD) sensor. A longpass filter is placed at the entrance of the spectrometer to suppress elastically scattered laser light. The sample is mounted on a motorized XY table and scanned under the microscope. Room-temperature PL measurements are performed using the same setup. For low-temperature PL measurements, the samples are mounted in vacuum on the cold finger of  a small He-flow cryostat, which can also be scanned under the microscope. In PL and Raman scanning experiments, full spectra are collected for sample positions defined on a square lattice. For all the scans performed on the CVD-grown sample, the 20x microscope objective was utilized. In order to extract information from these spectra, an automated fitting routine is employed, which yields the integrated intensity, spectral position and full width at half maximum (FWHM) of the characteristic PL and Raman spectral features. To study valley polarization effects, near-resonant excitation is employed in the PL setup. For this, a tunable, frequency-doubled pulsed fiber laser system is utilized. Its center wavelength is tuned to 633~nm. This laser is circularly polarized by a quarter-wave plate and coupled into the microscope system. A longpass filter is utilized to suppress scattered laser light, and the circular polarization of the PL is analyzed using a second quarter-wave plate and a linear polarizer placed in front of the spectrometer.
\section{Results and discussion}
\subsection{Photoluminescence}
First, we discuss the PL spectra of the CVD-grown films and compare them to those of exfoliated MoS$_2$. Pronounced PL is only observed in single-layer MoS$_2$ flakes, with a large reduction of the PL yield for two and more layers of MoS$_2$, and a corresponding redshift of the PL maximum peak position with increasing number of layers~\cite{Heinz_PRL10,Splen_Nano10}. The characteristic PL emission observed in single-layer MoS$_2$ at room temperature stems from the A exciton peak, which corresponds to a direct transition from the conduction band minimum to the uppermost valence band maximum at the K points in the Brillouin zone.

\begin{figure}[h]%
\centering
\includegraphics[width= \linewidth]{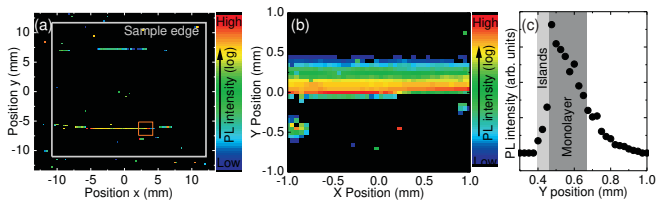}
\caption{(a) False color plot of the A exciton PL intensity as a function of position on the CVD-grown sample measured at room temperature. The  small orange square marks the region in which the higher-resolution scan shown in (b) was performed. (b) False-color plot of the A exciton PL intensity as a function of position. (c) High-resolution line scan of the A exciton PL intensity as a function of position along the vertical direction. PL data shown in (b) and (c) was collected at liquid-helium temperature.}
\label{4K_PL_maps}
\end{figure}

Figure~\ref{4K_PL_maps}(a) shows a false color plot of the A exciton PL intensity for the whole CVD-grown sample. The corresponding scanning PL measurement was performed using a step size of 250~$\mu$m. We note that strong PL is observed in two strip-like regions on the sample, which are about 8~mm long each. The width of these strips cannot be discerned directly due to the low resolution of the large-area scan shown in Fig.~\ref{4K_PL_maps}(a). To study these regions in more detail, high-resolution scans were performed in the area denoted by the orange square in Fig.~\ref{4K_PL_maps}(a). False color plots created from such a scan, with a scan region of 2~mm and a step size of 50~$\mu$m, are shown in Fig.~\ref{4K_PL_maps}(b). This scan was performed at liquid-helium temperature. From this scan, we can determine the width of the region showing strong PL to be on the order of 200~$\mu$m, in agreement with the width of the regions showing homogeneous contrast observed in the optical microscopy. We also note that the increase of the PL intensity is gradual at the upper boundary of this region (corresponding to the transition from a monolayer covered with additional material to a bare monolayer of MoS$_2$), while it drops sharply at the bottom (corresponding to the transition from continuous monolayer to isolated islands and bare substrate). This behavior is also seen in the higher-resolution line scan of the PL intensity shown in Fig.~\ref{4K_PL_maps}(c).  Thus, the areas of the CVD-grown samples which show PL comparable to single-layer MoS$_2$ flakes exceed the typical size of exfoliated flakes by several orders of magnitude, and  allow for optical spectroscopy experiments without the use of high-resolution microscope objectives.

\begin{figure}[h]%
\centering
\includegraphics[width= \linewidth]{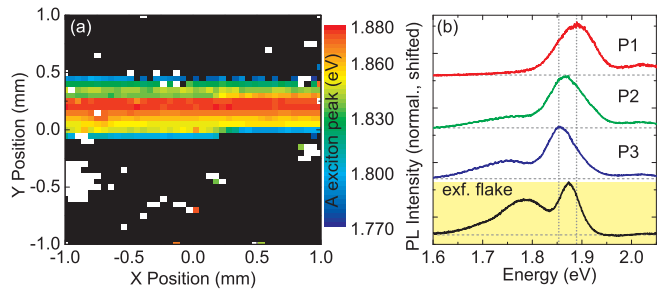}
\caption{(a) False color plot of the spectral position of the A exciton PL peak. The scan area is identical to that shown in Fig.~\ref{4K_PL_maps}(b). (b) Normalized PL spectra measured on different positions of the scan region shown in (a) and on an exfoliated flake at liquid-helium temperature.}
\label{PL_LowT}
\end{figure}

Now, we investigate the PL spectra in more detail. The PL data obtained in the scan shown in Fig.~\ref{4K_PL_maps}(b) allows us to extract the spectral position of the A exciton peak as a function of position on the sample. This is shown in Fig.~\ref{PL_LowT}(a) as a false color plot. We clearly see that the peak position shifts by more than 30~meV in the region which emits strong PL. This shift is pronounced in the direction perpendicular to the long axis of the region, while the peak position is nearly constant along the long axis. Individual spectra measured at different positions of the scan region are shown in Fig.~\ref{PL_LowT}(b), the positions P1-P3 are  separated by 100~$\mu$m along the $y$ direction. The vertical lines in Fig.~\ref{PL_LowT}(b) serve as a guide to the eye to mark maximum and minimum A exciton peak positions. The spectral shift indicates that the growth conditions, and the corresponding microscopic properties of the MoS$_2$ layer, change along the $y$ direction. Several effects may contribute to this large spectral shift. On the one hand, it was shown theoretically that the band gap of MoS$_2$ can be strongly influenced by the application of strain~\cite{Scalise12}, where biaxial compressive strain increases the band gap, while biaxial tensile strain decreases the band gap. Biaxial strain can be generated during the growth process and the subsequent cooling, as the MoS$_2$ and the SiO$_2$ substrate have very different thermal expansion coefficients~\cite{Plechinger12}. On the other hand, modifications of the dielectric environment modify the exciton binding energy by changing the Coulomb screening, and therefore shift the exciton emission peak. Such a change of the dielectric environment can be caused by additional material being deposited on top of the MoS$_2$ layer, e.g., an amorphous sulfur layer.   Additionally, in  n-doped MoS$_2$ layers, both, negatively charged excitons (trions) and neutral A excitons can be observed, with a binding energy of about 20~meV for the trions~\cite{Heinz_Trions}. These two emission peaks typically merge due to their large linewidths, and the peak position of the resulting PL emission then depends on the relative contributions by excitons and trions. Transport measurements on our CVD-grown films indicate n-type conductivity~\cite{Bartels13}, so that the observed PL emission may stem from a superposition of neutral and charged excitons, with a gradual change of the spectral weight due to changing carrier concentration.

In addition to the shift of the A exciton peak position, we also note that the FWHM of the peaks changes as a function of position, with values between 80~meV and 60~meV. Direct comparison of these PL spectra with a typical spectrum obtained from a single-layer MoS$_2$ flake at liquid-helium-temperatures demonstrates that the exfoliated MoS$_2$ has a significantly lower FWHM of about 37~meV. This indicates a larger inhomogeneous broadening of the A exciton transition in the CVD-grown sample due to local disorder, which cannot be resolved with the spatial resolution obtained in the PL experiment. In exfoliated MoS$_2$ flakes at low temperatures, we also observe a second, lower-energy PL peak (S exciton), which is associated with localized excitons bound to surface adsorbates~\cite{Korn_APL11,Plechinger12}. This peak is also observed in some areas of  the CVD-grown sample, e.g., at position P3 and weakly at position P2, yet absent on the majority of the regions emitting strong PL. A complete suppression of this S exciton peak was previously observed in experiments on oxide-covered MoS$_2$ flakes~\cite{Plechinger12}, its absence on most of the CVD-grown sample may therefore indicate partial coverage of the CVD monolayer with additional material.

\begin{figure}[h]%
\centering
\includegraphics[width=0.9 \linewidth]{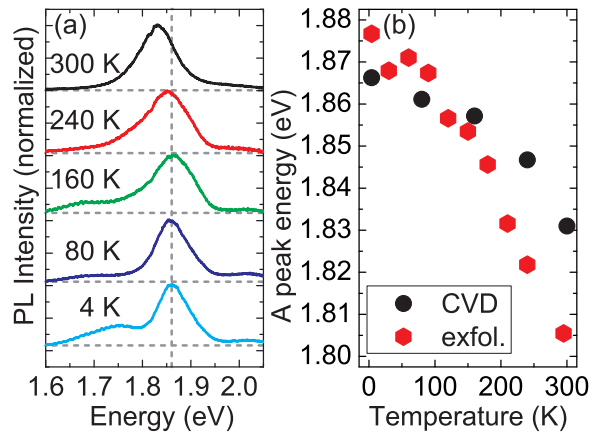}
\caption{(a) Normalized PL spectra measured as a function of sample temperature.  (b) A exciton peak position as a function of temperature for the CVD-grown sample (black dots) and an exfoliated MoS$_2$ flake (red hexagons).}
\label{T_Serie_2Panel}
\end{figure}

Next, we discuss the temperature dependence of the PL emission. Figure~\ref{T_Serie_2Panel}(a) shows a series of PL spectra measured on the CVD-grown sample for different sample temperatures. In order to avoid thermal drift of the sample position during changes of the temperature, a two-dimensional PL scan was performed at each temperature, and the spectra shown in Fig.~\ref{T_Serie_2Panel}(a) were created from these scans by averaging data in a 300~$\mu$m wide region. We note that the maximum of the A exciton emission redshifts by 35~meV  in the temperature range from 4~K to 300~K, as shown in Fig.~\ref{T_Serie_2Panel}(b). Additionally, the spectral width of the A exciton emission increases with increasing temperature. By contrast, in exfoliated MoS$_2$ flakes, we observe a spectral redshift of the A exciton peak by 72~meV in the same temperature range. The redshift indicates a temperature-induced reduction of the band gap due to thermal expansion of the crystal lattice. The fact that the redshift is far less pronounced in the CVD-grown sample shows that the CVD-grown MoS$_2$ film strongly adheres to the SiO$_2$ substrate, which has a very small thermal expansion coefficient. Similar effects were observed in oxide-covered MoS$_2$ flakes on SiO$_2$ substrates~\cite{Plechinger12}.

\begin{figure}[h]%
\centering
\includegraphics[width= \linewidth]{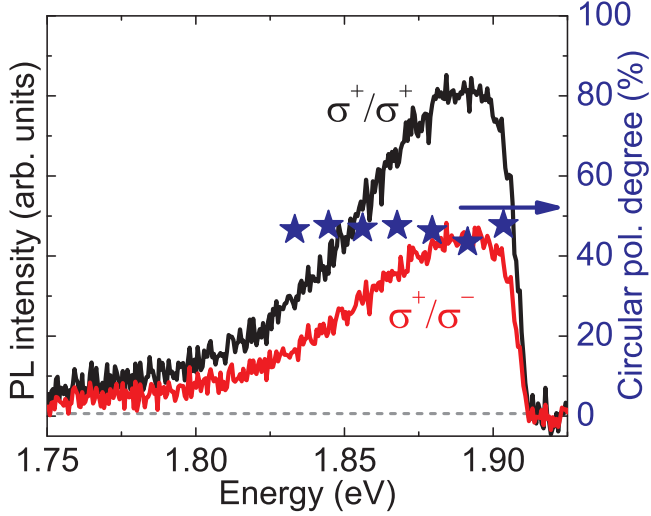}
\caption{Polarization-resolved PL spectra measured  at liquid-helium temperature under near-resonant, circularly-polarized excitation, for co- and contra-circular excitation/detection helicities. The star symbols indicate the circular polarization degree of the PL as a function of energy.}
\label{PL_circPol}
\end{figure}

Finally, we discuss the circular polarization of the PL of the CVD-grown sample at low temperatures. In monolayer MoS$_2$, the lack of an inversion center in the crystal structure leads to optical selection rules which allow for valley-selective excitation of electron-hole pairs either at the K$^+$ or the K$^-$ points in the Brillouin zone~\cite{Xiao12}, depending on the photon helicity. If this optically generated valley polarization persists on the timescale of the photocarrier recombination time, it is directly observable in the circular polarization degree of the PL~\cite{Cao_CircPol,Yao12,Heinz12,Marie_Valley12}. Figure~\ref{PL_circPol} shows two PL spectra collected for co- and contra-circular excitation and detection under near-resonant excitation using the frequency-doubled pulsed fiber laser system. In this experiment, the high-energy part of the A exciton emission is blocked by the longpass filter which is used to suppress the excitation laser. For co-circular excitation and detection, we note a significantly larger PL intensity than for contra-circular excitation and detection. From the two spectra, the circular polarization degree of the PL is calculated by dividing the difference of the PL intensities by their sum. For this, the PL intensities are spectrally averaged in  15~meV wide windows to reduce the noise. The circular polarization degree is plotted as a function of PL energy in Fig.~\ref{PL_circPol}.  We find a near-constant value of more than 40~percent in the entire spectral range of the A exciton emission that is observable. This value is comparable to values we observe in exfoliated monolayer MoS$_2$ flakes under similar experimental conditions (not shown). Recently, it was demonstrated that the circular polarization degree strongly depends on the excess energy provided by the excitation laser~\cite{kioseoglou:221907}.  The large circular polarization degree of the PL we observe is a clear indication that photocarrier excitation and recombination in the CVD-grown films follow the same selection rules as in monolayer MoS$_2$. Therefore, these films will be highly useful for future studies of coupled spin and valley physics in monolayer MoS$_2$.
\subsection{Raman spectroscopy}
Now, we discuss the Raman spectra of the CVD-grown films and compare them to those of exfoliated MoS$_2$. For MoS$_2$ prepared by mechanical exfoliation, Raman spectroscopy is an excellent tool to map the number of layers in individual flakes. The most prominent Raman modes that can be observed in MoS$_2$ under nonresonant excitation are the  E$^1_{2g}$ mode, which corresponds to an in-plane optical vibration of Mo and S atoms, and the A$_{1g}$, which is an out-of-plane optical vibration of the S atoms. The frequencies of both modes depend on the number of layers: while the A$_{1g}$ mode stiffens with additional layers, the E$^1_{2g}$ mode anomalously softens with additional layers~\cite{Heinz_ACSNano10}. This softening is attributed to enhanced dielectric screening~\cite{Molina11}. Thus, the difference $\omega_{Diff}= \omega_{A_{1g}}-\omega_{E^1_{2g}}$ between the two modes can be used as a fingerprint for the number of layers. This technique is well-suited to differentiate between monolayers and flakes of up to 4 layers, for thicker layers, the mode positions approach the bulk limit. Thicker flakes can be differentiated by analyzing the frequency of the Raman shear mode, which also shows a strong dependency on the number of layers~\cite{plechinger:101906, Zeng_ShearMos, Ferrari_ShearMos}.
\begin{figure}[h]%
\centering
\includegraphics[width= \linewidth]{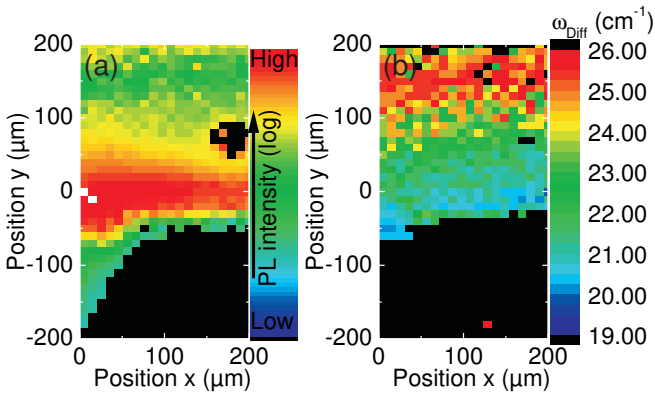}
\caption{(a) False color plot of the A exciton PL intensity as a function of position on the CVD-grown sample measured at room temperature. (b) False color plot of the frequency difference of the A$_{1g}$ and E$^1_{2g}$ Raman modes as a function of position for the same scan area as shown in (a).}
\label{2panel_Raman_PL_maps}
\end{figure}
\begin{figure}[h]%
\centering
\includegraphics[width= \linewidth]{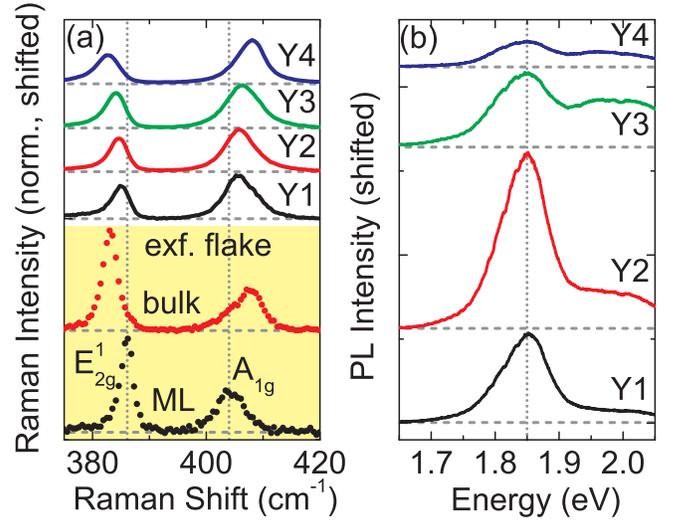}
\caption{(a) Raman spectra measured on different positions of the CVD-grown film and on  exfoliated monolayer (ML) and bulk-like flakes. All spectra are normalized to the amplitude of the A$_{1g}$ mode. The vertical lines mark the positions of A$_{1g}$ and E$^1_{2g}$ in an exfoliated monolayer and serve as guide to the eye.  (b) PL spectra measured on the same positions of the CVD-grown film as shown in (a).}
\label{Raman_PL_2panel}
\end{figure}

In order to determine the information that can be extracted from Raman spectra on the CVD-grown films, we performed combined Raman and PL scans at room temperature, so that for each spatial position, both, Raman and PL spectra were collected. False color plots created from these scans are depicted in Fig.~\ref{2panel_Raman_PL_maps}. Here, the A exciton intensity is plotted in Fig.~\ref{2panel_Raman_PL_maps}(a), this map is directly comparable to the frequency difference $\omega_{Diff}$ plotted in Fig.~\ref{2panel_Raman_PL_maps}(b). Similar to the high-resolution PL scan shown in Fig.~\ref{4K_PL_maps}(a), we observe a rapid increase of the PL intensity along the vertical direction corresponding to the transition from bare substrate to islands and continuous film. In the corresponding frequency difference map, we see a gradual increase of $\omega_{Diff}$ along the vertical direction, with  less well-defined contrast than in the PL intensity map.

We now discuss the Raman spectra in more detail.  Fig.~\ref{Raman_PL_2panel}(a) shows 4 Raman spectra collected at different positions on the CVD film, the corresponding PL spectra are depicted in Fig.~\ref{Raman_PL_2panel}(b). Here, the position Y1 corresponds to the film region where individual islands are observed in the optical microscope, while Y2-Y4 are positions on different parts of the monolayer-like region.  Let us first compare the Raman spectra to those collected from exfoliated flakes. All spectra shown in Fig.~\ref{Raman_PL_2panel}(a) are normalized to the A$_{1g}$ mode amplitude. We clearly see that in the CVD-grown sample, the  E$^1_{2g}$ amplitude is smaller than the  A$_{1g}$, while in the exfoliated flakes, the opposite is observed. Additionally, we note that the linewidths for both Raman modes in the CVD-grown film are larger than in the exfoliated flake, and the E$^1_{2g}$ mode is asymmetric. The mode positions for the CVD-grown film are in between the values observed for monolayer and bulk-like exfoliated flakes. A  small increase of the  frequency difference $\omega_{Diff}$ (less than 0.5~cm$^{-1}$) is observed for the transition from isolated islands (Y1) to the film region showing the largest PL signal (Y2). The regions which emit less PL show further increase of the  frequency differences, with position Y4 having a value of $\omega_{Diff} = 25$~cm$^{-1}$, comparable to bulk-like exfoliated MoS$_2$. For exfoliated MoS$_2$, however, a drastic reduction of the PL emission intensity, accompanied by a large redshift of the PL peak position, is observed for the change from monolayer to bilayer~\cite{Heinz_PRL10,Splen_Nano10} and thicker flakes. From the different evolution of $\omega_{Diff}$ and PL intensity we observe in the CVD-grown film, we infer that the Raman frequency shift is not caused by the formation of well-defined bilayer and multilayer regions, which would quench the PL due to a change of the band structure to an indirect band gap. By contrast, the monolayer may partially be covered by additional material which influences the dielectric screening and the vibronic properties, yet does not drastically change the band structure.
\section{Conclusion}
In conclusion, we have used Raman and photoluminescence spectroscopy to directly compare CVD-grown MoS$_2$  to exfoliated flakes. The CVD process used to prepare our films yields large areas which show PL comparable to monolayer MoS$_2$ flakes. In low-temperature PL measurements, we observe a spectral shift of the A exciton peak as a function of position. The linewidth of the A exciton is larger than that observed in exfoliated MoS$_2$. Some regions of the CVD-grown film also show PL emission associated with surface-adsorbate-bound excitons. Temperature-dependent PL measurements reveal a small redshift of the A exciton peak, indicating that the CVD-grown MoS$_2$ strongly adheres to the SiO$_2$ substrate. In low-temperature PL experiments under near-resonant excitation, we observe a large circular polarization degree of the PL, commensurate with an optically induced valley polarization. In scanning Raman experiments, we observe a substantial shift of the characteristic Raman modes of MoS$_2$ in the region of the CVD-grown film which emits strong PL. This behavior differs from the Raman spectra observed in exfoliated MoS$_2$, where similar shifts only occur for an increasing number of layers and a corresponding suppression of the PL due to changes in the band structure.
Our observations indicate that the CVD-grown MoS$_2$ material is well-suited for future studies of exciton physics and valley effects. Further optimization of the growth process should yield even larger, homogeneous MoS$_2$ regions and may pave the way for device applications.
The authors  acknowledge financial support by the DFG via SFB689, KO3612/1-1 and GRK 1570. L.B. gratefully acknowledges support by C-SPIN, one of six centers supported by the STARnet phase of the Focus Center Research Program (FCRP), a Semiconductor Research Corporation program sponsored by MARCO and DARPA. Additional support originates from the US National Science Foundation under grants DMR-1106210.
\section*{References}
\bibliography{MoS2}
\end{document}